\documentclass{PoS}

\title{Nuclear Dependence of Transverse Single-Spin Asymmetries in Polarized $p$+$A$ Collisions at RHIC}

\ShortTitle{Nuclear Dependence of Transverse Single-Spin Asymmetries in Polarized $p$+$A$ Collisions at RHIC}

\author{\speaker{Stephen Pate} [for the PHENIX Collaboration]\\
        Physics Department, New Mexico State University, Las Cruces NM 88003, USA\\
        E-mail: \email{pate@nmsu.edu}}

\abstract{Large transverse single-spin asymmetries (TSSA) in hadron production at forward rapidity have 
been observed in polarized $p$+$p$ interactions for many decades, over a large range of center-of-mass 
energies, and have led to the investigation of spin-momentum correlations such as the Sivers and Collins 
effects.  In the last few years, it has been discovered at RHIC that these single-spin asymmetries may 
be enhanced or suppressed in $p$+$A$ collisions, and the nuclear-size and centrality dependence have 
been studied.  A variety of phenomena have been observed and likely they do not have all a single 
explanation; we see an apparent quenching of the TSSA in forward charged hadron production with 
increasing nuclear size, while we see an enhancement in the asymmetry in $J/\psi$ production, and 
in very forward neutron production even a sign change in the asymmetry is seen.  Other systems, 
such as $\pi^0$ production at central rapidity, do not display a nuclear-size dependence.  These 
observations provide a bridge between the study of the initial state in heavy-ion collisions and 
that of the nucleon spin puzzle, and open up a new method for the investigation of cold nuclear matter.}

\FullConference{23rd International Spin Physics Symposium - SPIN2018 -\\
		10-14 September, 2018\\
		Ferrara, Italy}

\begin{document}

\bibliographystyle{siam}

\section{Transverse Single Spin Asymmetries in $\vec{p}+p$ Interactions}

Large transverse single-spin asymmetries (TSSAs) 
have been observed in forward hadron production in $\vec{p}+p$ interactions
for many decades and continue to be a subject of experimental and theoretical investigation.  (And they are 
the topic of many talks at this conference!)   A major result of the multi-decades study of this phenomenon is
an understanding of the importance of various spin-momentum correlations.
\begin{itemize}
\item Initial state correlations:  For example, the Sivers effect correlates the transverse spin of the
nucleon with the intrinsic transverse momentum of the resident quark \cite{PhysRevD.41.83}.
\item Final state correlations: For example, the Collins effect correlates the transverse spin of the
fragmenting quark with the transverse momentum of the produced hadron \cite{COLLINS1993161}.
\end{itemize}
Direct measurements of these functions in semi-inclusive deep-inelastic scattering and in 
$e^+e^-$ annihilation interactions can be applied back to $\vec{p}+p$ interactions to reproduce data with
some success \cite{PhysRevD.87.094019}.  
We have learned that fragmentation (i.e.\ the Collins effect) seems to be the main culprit
for large asymmetries see in $\vec{p}+p\rightarrow\pi +X$~\cite{PhysRevD.89.111501}, 
at least in the $x_F<0.3$ region; for
$x_F>0.3$ other mechanisms must be invoked as well \cite{PhysRevD.86.074032}.  An excellent overview of this field
can be found in the review article by Pitonyak \cite{Pitonyak:2016hqh}.

The main focus of this presentation is:  What happens when these interactions occur in a nuclear
environment?

\section{Transverse Single Spin Asymmetries in $\vec{p}+A$ Interactions}

The nuclear environment brings many additional physics processes into play with many possible outcomes for 
these asymmetries.  We have been aware since the 1980s of the changes the nucleus can have on the observed
nucleon parton distribution functions (the EMC effect and nuclear shadowing), which represents a modification
of the initial state.  Gluon saturation and/or the existence of a Color Glass Condensate may also modify
the magnitude of TSSAs.  Multiple rescatterings of the final state particles may result in a broadening of the
transverse distribution functions (leading to a dilution of the asymmetry) or to a modification of the effective
$x_F$ (which might enhance the asymmetry).  Ultra-peripheral collisions can induce very strong electromagnetic
fields and Primakoff-like effects.

An important issue that arises in the analysis of $p+A$  interactions is the centrality of the interaction; 
how closely did the proton pass near the center of the nucleus?  The impact parameter determines the average number
of binary $p+p$ and $p+n$ collisions ($N^{\rm Avg}_{\rm coll}$) that will occur as the proton 
passes through the nucleus.  Of course the impact parameter cannot be measured directly; instead we observe the 
the charged-particle multiplicity at extreme rapidity in a set of ``Beam-Beam Counters'' and use this to
establish a ``centrality classification'' from 0\% to 100\%.  A model calculation then 
determines $N^{\rm Avg}_{\rm coll}$.

\section{The PHENIX Experimental Apparatus}

\begin{figure}
\begin{center}
\includegraphics[scale=0.5]{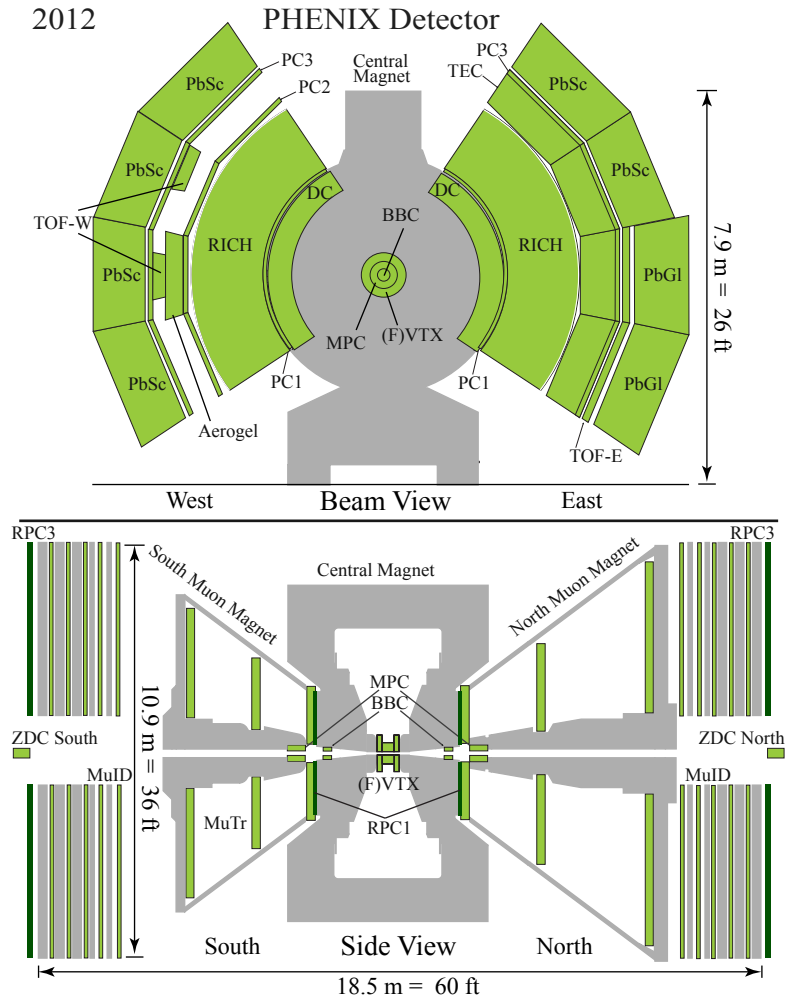}
\end{center}
\caption{Overview of the PHENIX apparatus; upper panel shows the central arms, while the lower panel
shows the muon arms, the Beam-Beam Counters (BBCs), and the Zero Degree Calorimeters (ZDCs).}
\label{PHENIX}
\end{figure}

PHENIX operated as a collider experiment at the Relativistic Heavy-Ion Collider (RHIC) at Brookhaven 
National Laboratory (BNL) from 2001 to 2016.  It comprised two major detector systems:
\begin{itemize}
\item a pair of central arms covering
the rapidity range $|y|<0.35$ including vertex detectors, a bending magnet, tracking detectors, 
particle identification and calorimetry (upper panel of Figure~\ref{PHENIX}) 
\item a pair of muon arms covering the forward and backward rapidity range $1.4 < |y| < 2.4$ including vertex
detectors,
muon range shielding, muon tracking, a radial magnetic field, and a muon/hadron identifier (lower panel of
Figure~\ref{PHENIX})
\end{itemize}
In addition, at very forward rapidity a pair of Zero Degree Calorimeters (ZDCs), located beyond the accelerator
bending magnets (see the lower panel of
Figure~\ref{PHENIX}), detected neutrons emitted from the collisions.

The experimental data comes from the 2015 running period, a unique situation in which a 
transversely polarized proton
beam collided with either another beam of transversely polarized protons or a beam of unpolarized Al or Au nuclei,
with a nucleon-nucleon center-of-mass energy of $\sqrt{s_{NN}}=200$~GeV.

\section{TSSA for $\pi^0$ Production at Central Rapidity}

\begin{figure}
\begin{center}
\includegraphics[scale=0.4]{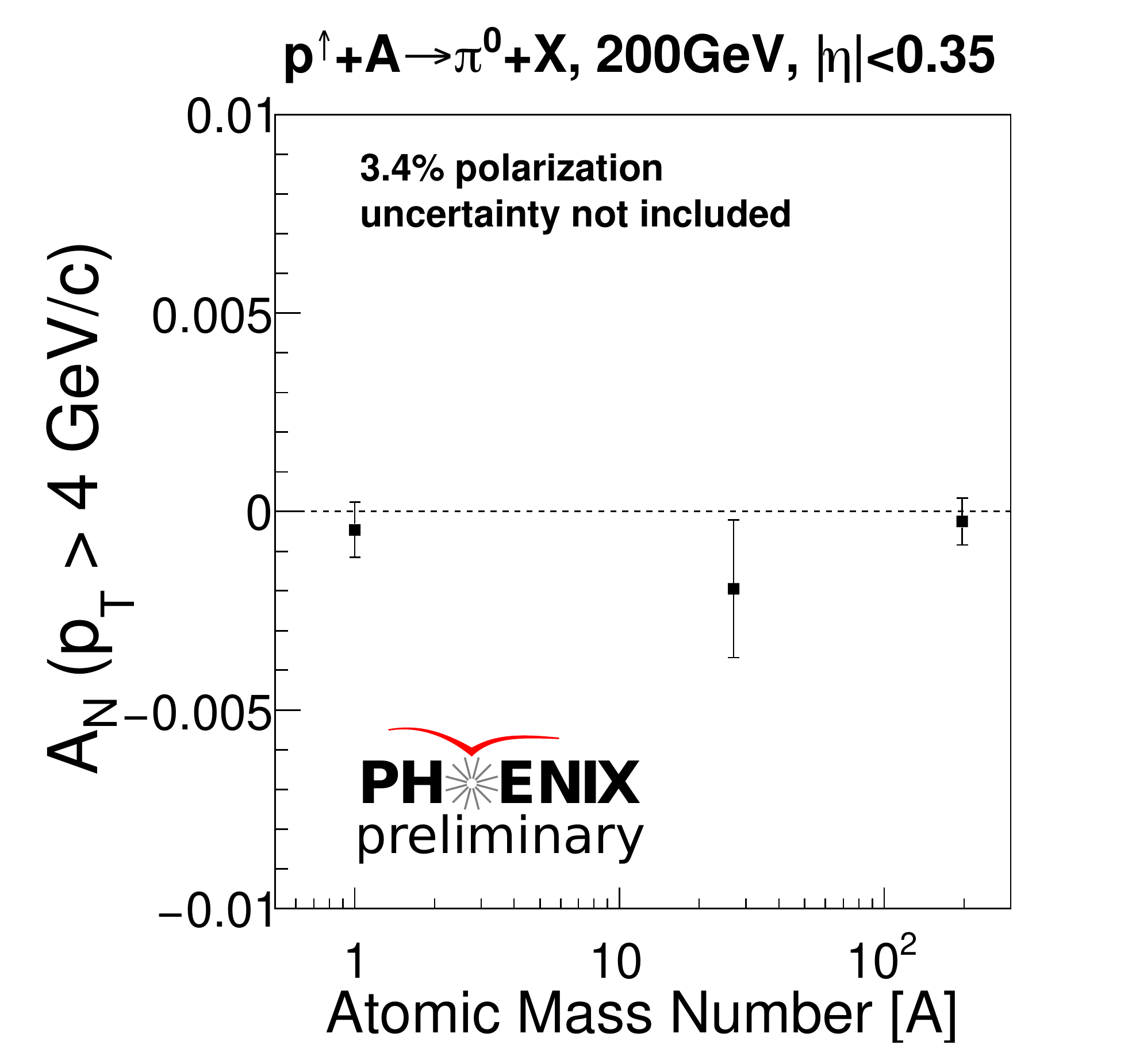}
\end{center}
\caption{The A-dependence of the transverse single spin asymmetry in $\pi^0$ production.}
\label{AN_pi0}
\end{figure}

This study used the tracking and calorimetric abilities of the central arms to reconstruct $\pi^0$ particles
from pairs of electromagnetic showers not associated with any charged particle in the trackers.  The TSSA for
$\pi^0$ production in $\vec{p}+p$, $\vec{p}+$Al, and $\vec{p}+$Au collisions is consistent with zero
over the transverse momentum range $1<p_T<16$ GeV.  Integrating over that $p_T$ range, we display
the A-dependence of the asymmetry in Figure~\ref{AN_pi0}, and the lack of any A-dependence at
central rapidity is very clear.  A manuscript describing this measurement is in preparation.

\section{TSSA for $J/\psi$ Production at Forward and Backward Rapidity}

\begin{figure}
\begin{center}
\includegraphics[scale=0.5]{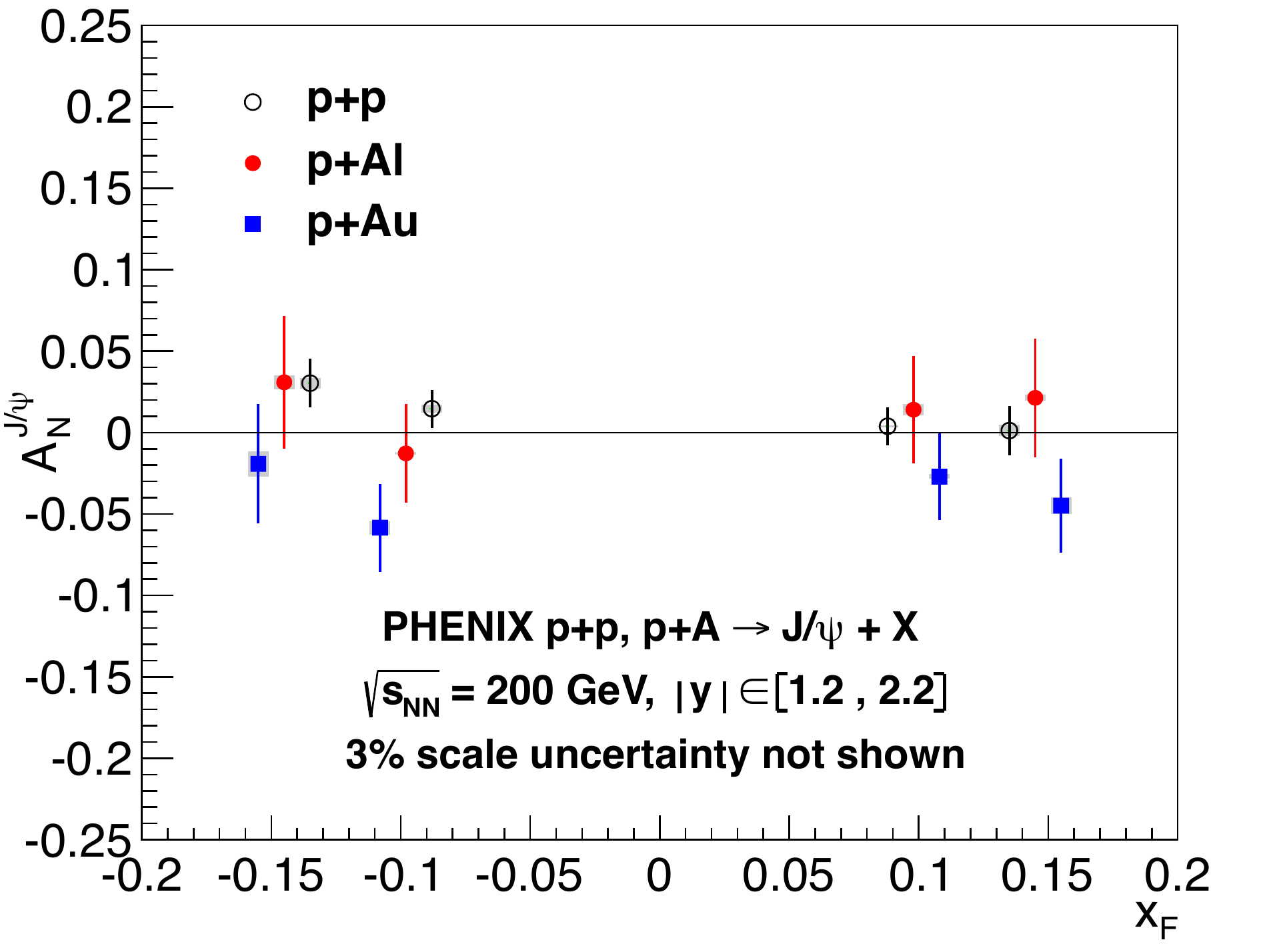}
\end{center}
\caption{The nuclear dependence of the TSSA in $J/\psi$ production, as a function of $x_F$.}
\label{AN_JPSI_XF}
\end{figure}

\begin{figure}
\begin{center}
\includegraphics[scale=0.6]{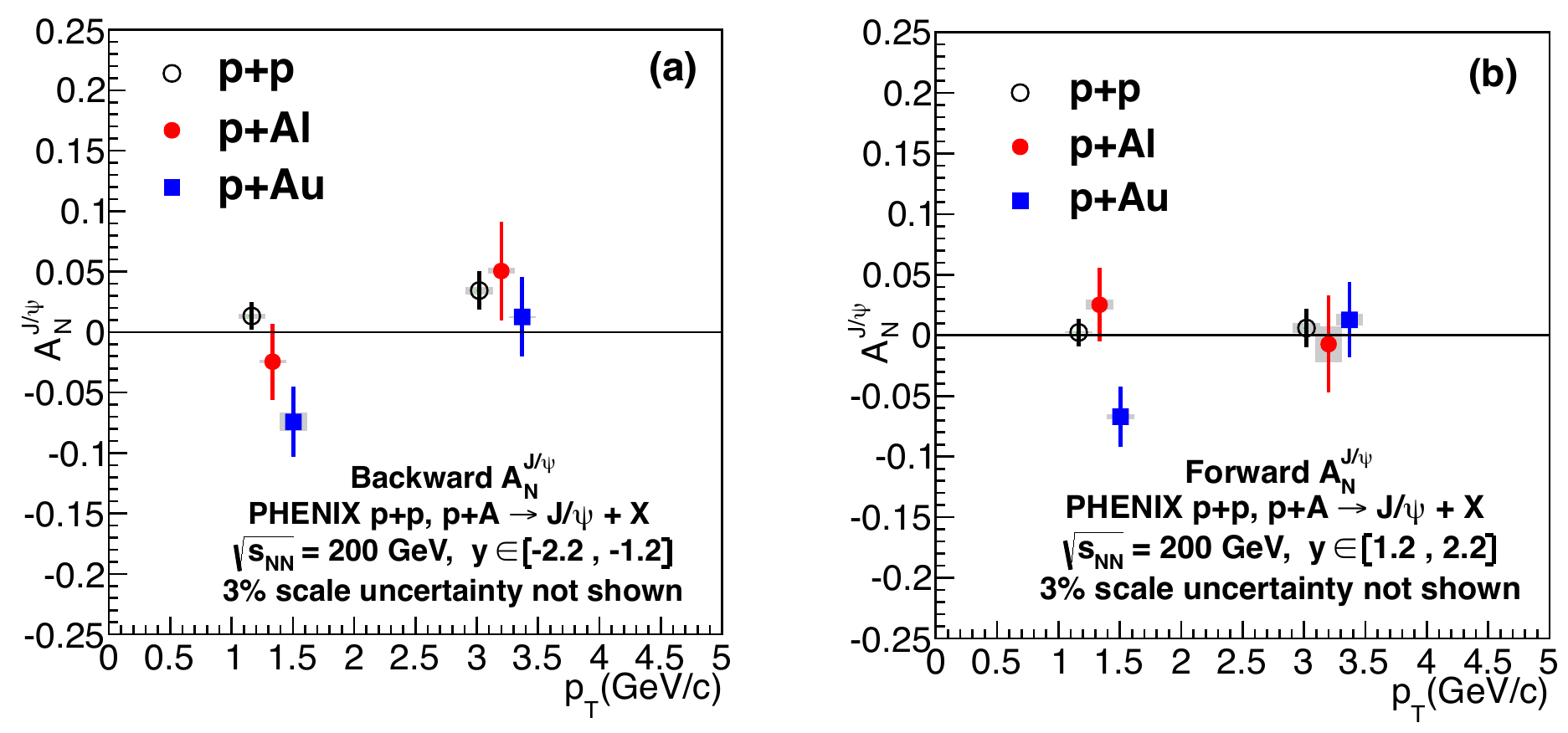}
\end{center}
\caption{The nuclear dependence of the TSSA in $J/\psi$ production, as a function of $p_T$.}
\label{AN_JPSI_PT}
\end{figure}

The TSSA for $J/\psi$ production in $\vec{p}+p$ collisions at forward and backward rapidity 
was previously measured in PHENIX and was seen to be consistent with
zero; in the 2015 run we were able to observe if this asymmetry had any nuclear dependence. We observe a 
negative asymmetry in $\vec{p}+{\rm Au}$ collisions in both forward and backward production; the
largest signal is seen for $-0.11 < x_F < -0.05$, as is seen in Figure~\ref{AN_JPSI_XF}.  Looking at
the results instead as a function of $p_T$, Figure~\ref{AN_JPSI_PT}, we observe a negative asymmetry 
in $\vec{p}+{\rm Au}$ collisions in
both forward and backward production for $0.42 < p_T < 2$ GeV.  
This work was published in \cite{PhysRevD.98.012006}.

It is evident that the nuclear environment has
created or enhanced the asymmetry.
This could be a new avenue for insight into the $J/\psi$ production mechanism in
nuclei, critical for an understanding of $J/\psi$ production modification in $A+A$
collisions.

\section{TSSA for $\pi^+/K^+$ Production at Forward Rapidity}

The asymmetry in the production of charged pions at forward rapidity, first seen in experiments 
at Argonne National Laboratory~\cite{PhysRevLett.36.929} and also seen much more recently 
at RHIC by BRAHMS~\cite{PhysRevLett.101.042001}, was the asymmetry that started the investigation of
TSSAs 40 years ago.  With the 2015 data at RHIC we had the opportunity to study the nuclear dependence
of this very fundamental production channel.  This study in PHENIX used the same muon arms as the study
of $J/\psi$ production described above, but used the Muon Identifier arms to {\em exclude} muon tracks,
thus providing a sample of unidentified hadron tracks composed almost entirely of pions and kaons.

\begin{figure}
\begin{center}
\includegraphics[scale=0.6]{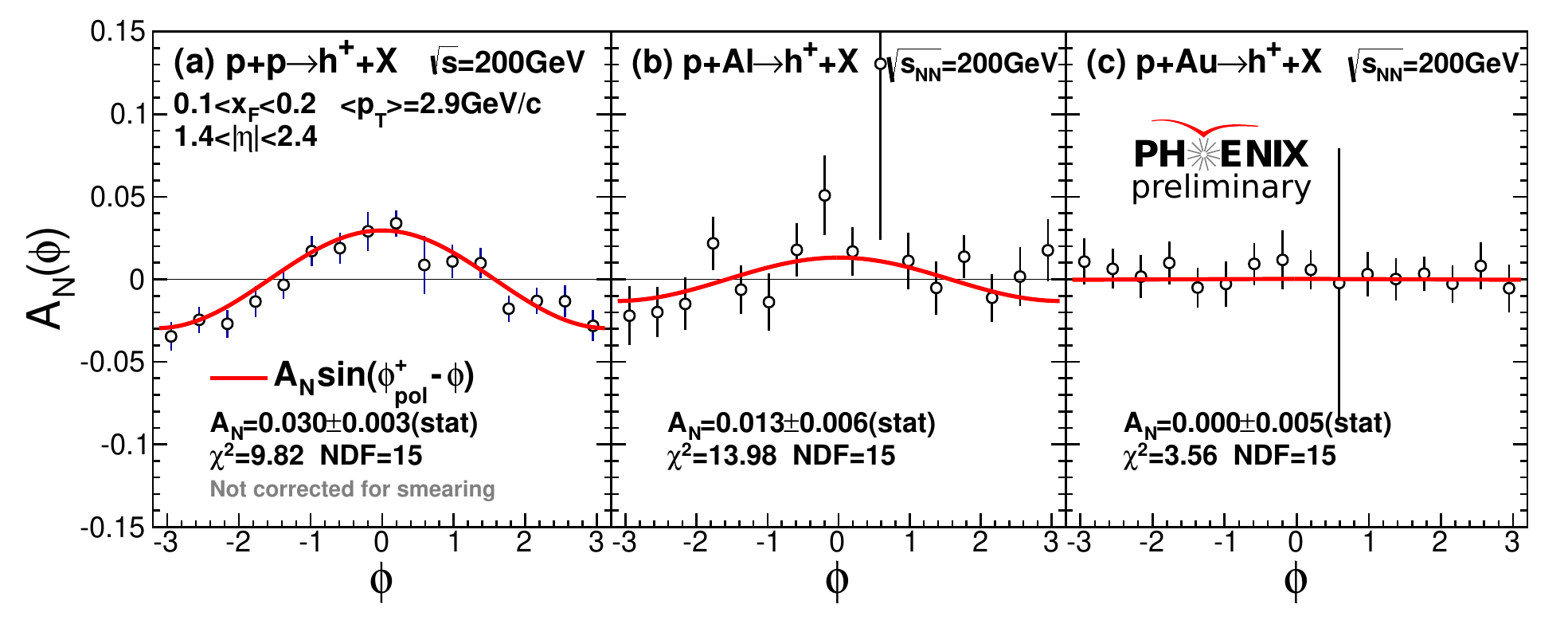}
\end{center}
\caption{The nuclear dependence of the TSSA in $\pi^+/K^+$ production.  Each panel displays a sinusoidal fit of
the azimuthal asymmetry of the yield, for $\vec{p}+p$, $\vec{p}+$Al, and $\vec{p}+$Au interactions.}
\label{AN_hadrons_fits}
\end{figure}

\begin{figure}
\begin{center}
\includegraphics[scale=0.6]{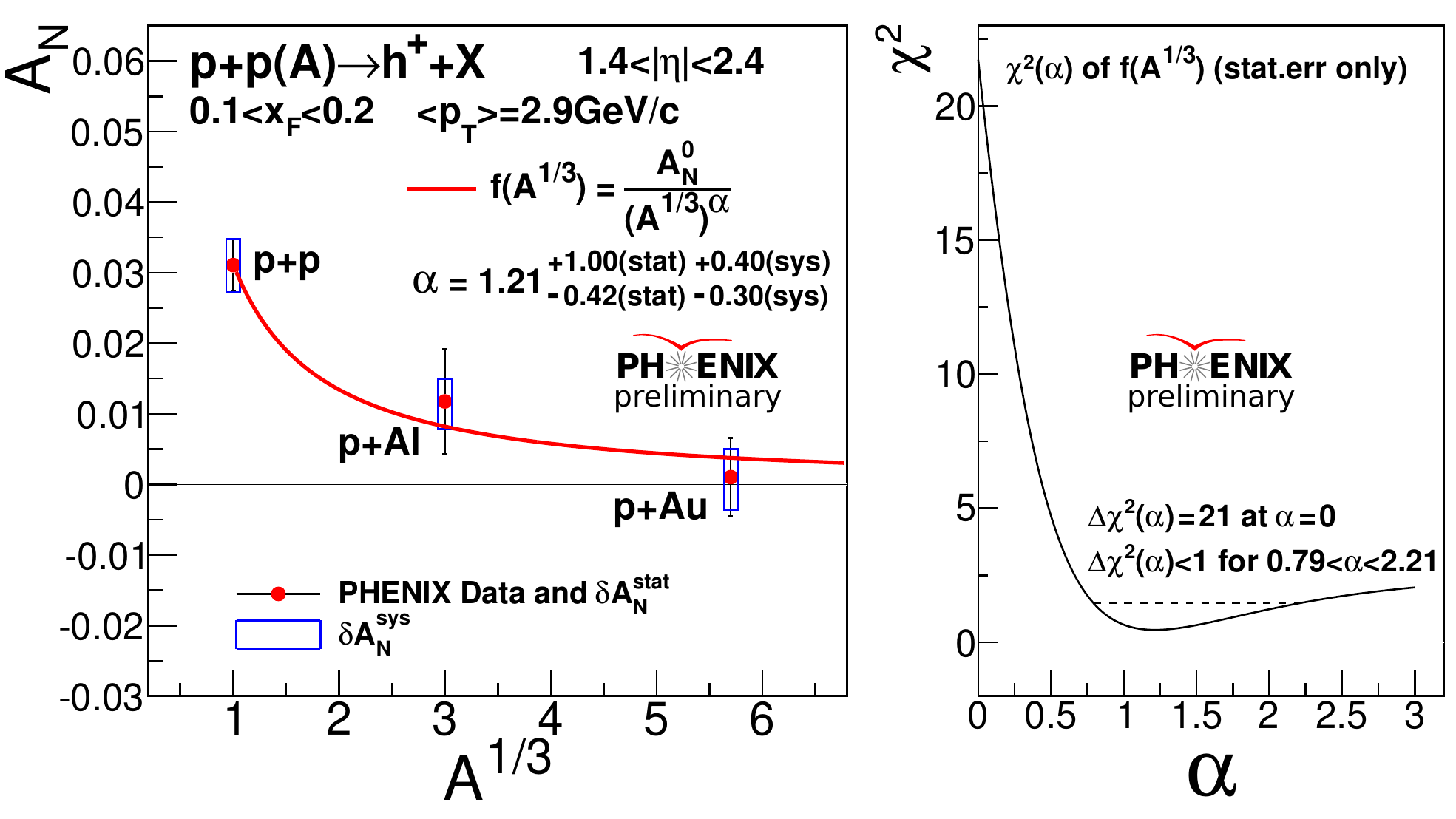}
\end{center}
\caption{Demonstration of the A$^{1/3}$-suppression of the TSSA in $\pi^+/K^+$ production.}
\label{AN_hadrons_mass}
\end{figure}

\begin{figure}
\begin{center}
\includegraphics[scale=0.6]{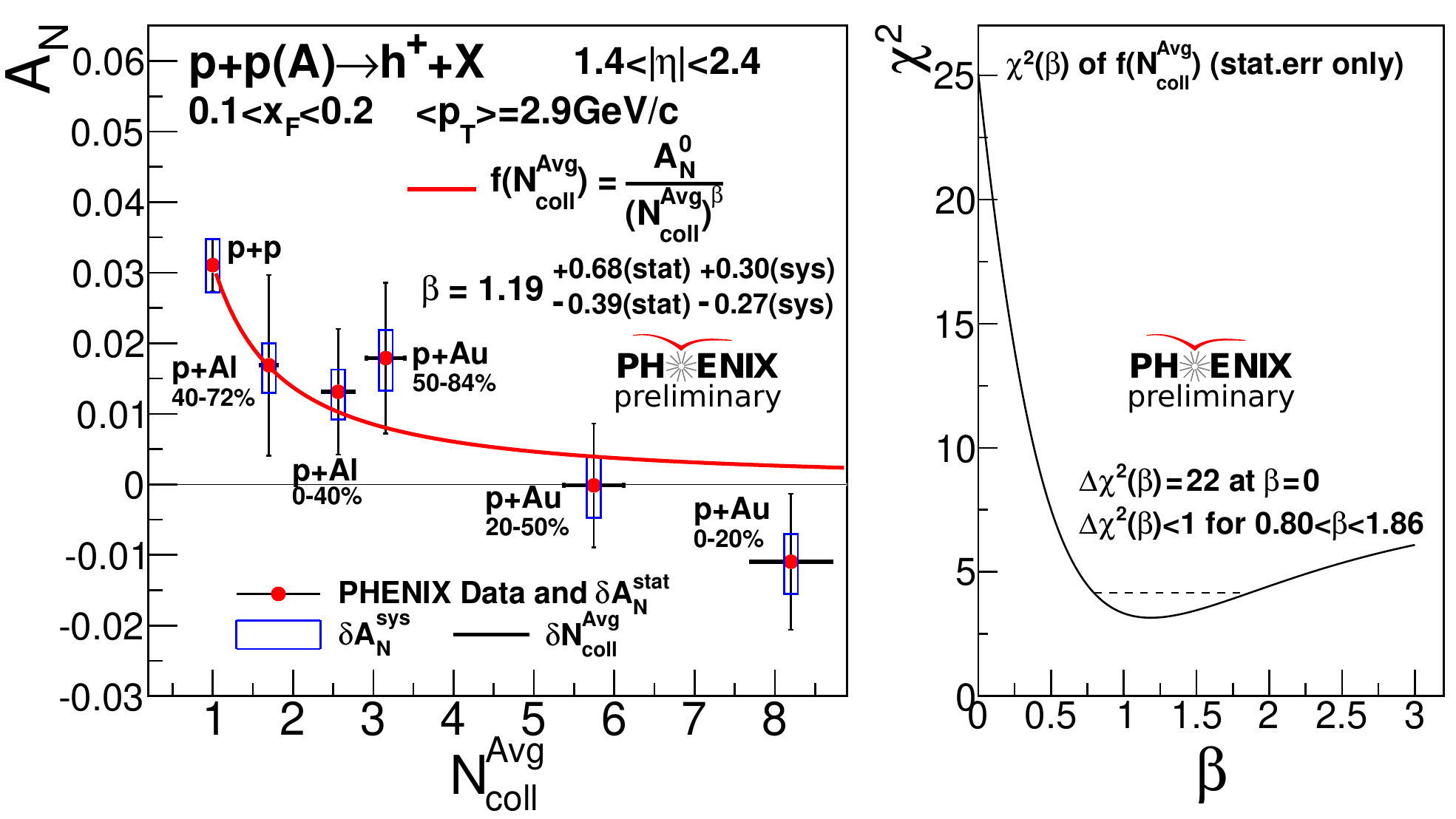}
\end{center}
\caption{Demonstration of the $N^{\rm Avg}_{\rm coll}$-suppression of the TSSA in $\pi^+/K^+$ production.
The $p+$Al and $p+$Au data points are labeled with the associated centrality class, given as a range of 
percentages; the most central collisions have the lowest centrality percentage.}
\label{AN_hadrons_cent}
\end{figure}

In Figure~\ref{AN_hadrons_fits} are shown sinusoidal fits to the azimuthal distribution of the spin
asymmetry.  A very strong suppression of the azimuthal
asymmetry is seen for increasing nuclear mass number.  In Figure~\ref{AN_hadrons_mass} we show a 
fit of the asymmetries to a function of the form A$^{-\alpha/3}$ so as to quantitatively describe the nuclear
dependence.  The best fit is for $\alpha = 1.21^{+1.08}_{-0.52}$.  This is consistent with the expectation
of Hatta et al.~\cite{PhysRevD.95.014008} that there should be an A$^{1/3}$ suppression of the final-state
Collins asymmetry due to gluon saturation.  Note that the possibility of ``no mass dependence'' ($\alpha=0$) is
ruled out.  Finally in Figure~\ref{AN_hadrons_cent} we display a fit of the asymmetry, binned into centrality
classes, as a function of the average number of binary collisions $N^{\rm Avg}_{\rm coll}$ calculated for each
centrality class range.  The fit is to a function of the form $(N^{\rm Avg}_{\rm coll})^{-\beta}$, and
the best fit gives $\beta = 1.19^{+0.74}_{-0.47}$.  Once again, the possibility of ``no mass dependence'' 
($\beta=0$ in this case) is ruled out.  A manuscript describing this measurement is in preparation.

\section{TSSA for Neutron Production at Very Forward Rapidity}

\begin{figure}
\begin{center}
\includegraphics[scale=0.5]{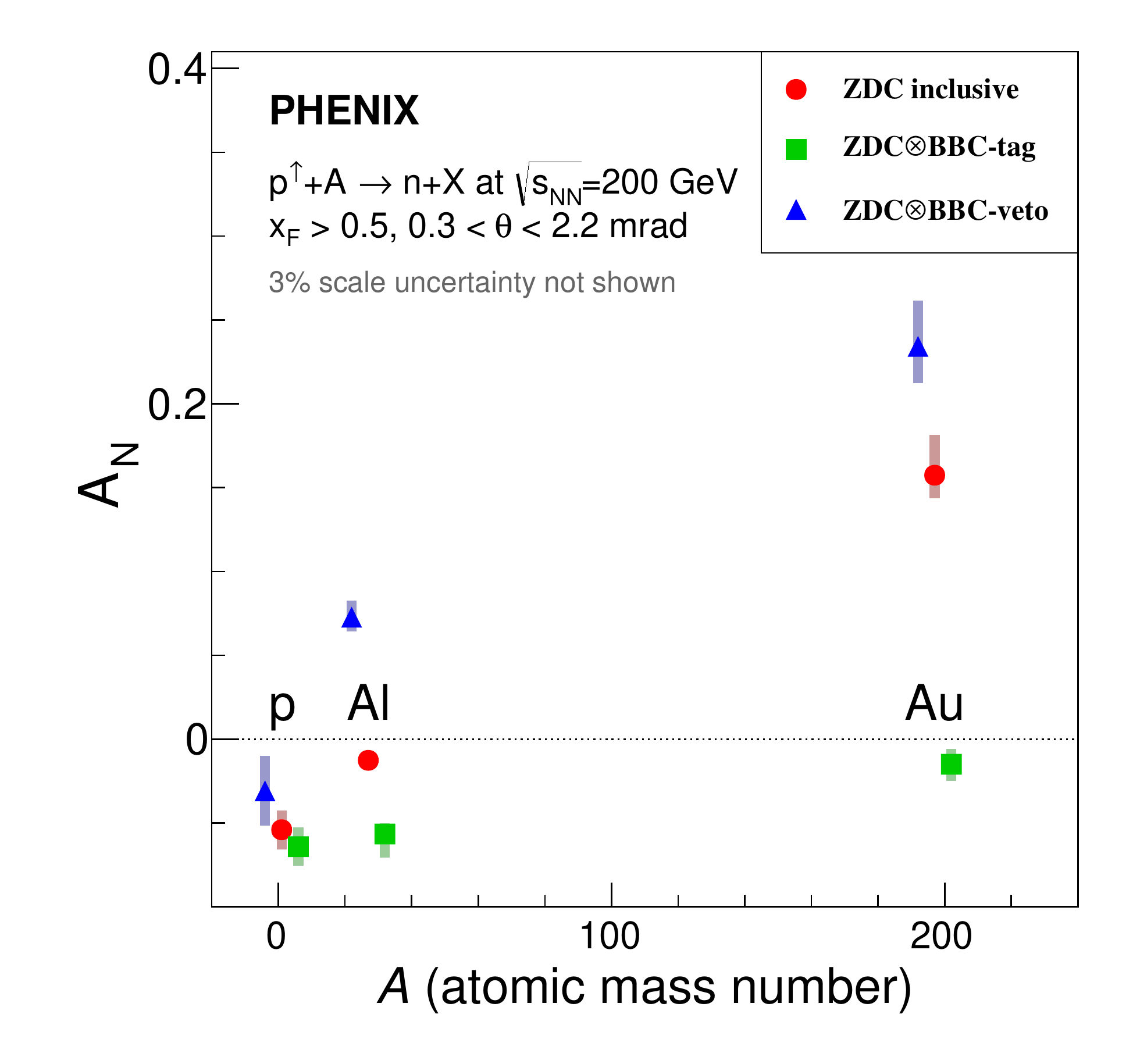}
\end{center}
\caption{Nuclear- and centrality-dependence of the TSSA in neutron production at very forward rapidity.}
\label{AN_n}
\end{figure}

The discovery at RHIC some years ago of the large TSSA for very forward neutron 
production in $\vec{p}+p$ collisions \cite{Bazilevsky:2003bm,PhysRevD.88.032006} was already a surprising result; the RHIC 2015 data
presented an opportunity to see if this asymmetry had a nuclear dependence.  The principal detector in use
for this study was the ZDC (zero-degree calorimeters) mentioned earlier; they are positioned beyond the 
bending magnets used to steer the beams into collision and so are sensitive primarily to neutrons arising from the
interaction point. In addition, the BBC (beam-beam counters) were used with the ZDC to create three different trigger conditions, to study the centrality dependence.
\begin{enumerate}
\item {\bf ZDC Inclusive:} ZDC energy $>$ 15 GeV (all collisions with significant ZDC energy)
\item {\bf ZDC$\otimes$BBC-tag:}  ZDC Inclusive AND $\ge 1$ hit in each BBC (hard/central collisions)
\item {\bf ZDC$\otimes$BBC-veto:}  ZDC Inclusive AND no hit in any BBC (diffractive/peripheral collisions)
\end{enumerate}
The results for each of these triggers are shown in Figure~\ref{AN_n}.  Collisions with small impact parameter
(BBC-tag) show a negative asymmetry with a moderate nuclear dependence.  On the other hand, collisions with
a large impact parameter (BBC-veto) show a strong nuclear dependence, indicating the importance of ultra-periphercal collisions.  Mitsuka~\cite{PhysRevC.95.044908} has shown that in an ultra-peripheral collision the photon
flux from the unpolarized nucleus can interact with the polarized proton to produce such asymmetries.  These
measurements have been published in \cite{PhysRevLett.120.022001}.

\section{Summary of Nuclear Dependence of TSSAs}

The RHIC 2015 data with $\vec{p}+p$, $\vec{p}+$Al, and $\vec{p}+$Au collisions has been used by PHENIX
to study the transverse single spin asymmetries in a number of systems.
\begin{itemize}
\item $\pi^0$ production at central rapidity: the TSSA is very small, independent of nuclear size
\item $J/\psi$ production at forward and backward rapidity: the TSSA is very small in $\vec{p}+p$, becoming negative
at low $p_T$ in $\vec{p}+$Au
\item $\pi^+/K^+$ production at forward rapidity:  the TSSA is very large in $\vec{p}+p$, and is then strongly
suppressed with increasing nuclear size and centrality, suggesting a gluon saturation effect
\item neutron production at very forward rapidity:  largest effect seen in ultra-peripheral collisions
\end{itemize}
A rich variety of behavior is revealed in this cross between spin-dependent scattering phenomena and the physics
of cold nuclear matter.

\section{Acknowledgements}
This work was supported by a grant from the US Department of Energy, Office of Science, 
Medium Energy Nuclear Physics program.

\bibliography{Pate_SPIN2018}

\end{document}